\documentclass{edm_article}

\usepackage{graphicx}
\usepackage{booktabs}
\usepackage{url}
\usepackage{microtype}
\begin{document}

\title{Comparing RAG and GraphRAG for Page-Level Retrieval Question Answering on a Math Textbook}

\numberofauthors{6}
\author{
Eason Chen\\
       \affaddr{Carnegie Mellon University}\\
       \email{eason.tw.chen@gmail.com}
\and
Chuangji Li\\
       \affaddr{Carnegie Mellon University}\\
       \email{chuangjl@andrew.cmu.edu}
\and
Eric Li\\
       \affaddr{Carnegie Mellon University}\\
       \email{shizhuol@andrew.cmu.edu}
\and
Zimo Xiao\\
       \affaddr{Carnegie Mellon University}\\
       \email{zimox@andrew.cmu.edu}
\and
Jionghao Lin\\
       \affaddr{The University of Hong Kong}\\
       \email{jionghao@hku.hk}
\and
Ken Koedinger\\
       \affaddr{Carnegie Mellon University}\\
       \email{koedinger@cmu.edu}
}
\maketitle

\begin{abstract}
Large language models (LLMs) show promise as educational aids but often lack alignment with specific course materials. We investigate Retrieval-Augmented Generation (RAG) and GraphRAG for page-level question answering on an undergraduate mathematics textbook. Using a curated dataset of 477 question-answer pairs, each tied to a specific textbook page, we compare five embedding-based RAG models, a BM25 baseline, and GraphRAG across two metrics: retrieval accuracy (whether the correct page is retrieved) and answer quality (F1 score). Our results show that embedding-based RAG outperforms GraphRAG for page-level retrieval, with \texttt{voyage-3-large} achieving 99.4\% accuracy at top-10 (bootstrap 95\% CI for top-1: [.644, .728]). BM25 proves a strong baseline, outperforming several embedding models. Error analysis reveals that 63.3\% of top-1 failures retrieve same-chapter content, suggesting pedagogical relevance even in failure cases. GraphRAG retrieves excessive context ($\sim$47K tokens vs.\ $\sim$3.7K for RAG), reducing generation quality. We further replicate key experiments using an open-source local LLM (Qwen3.5-35B-A3B), finding that RAG benefits are proportionally larger for weaker models (+39\% vs.\ +16\% relative F1 improvement), an important result for cost-sensitive educational deployments. These findings inform the design of AI tutoring systems that reference specific textbook pages.
\end{abstract}


\section{Introduction}

During self-paced study, students frequently need to locate specific textbook pages, for example to revisit a definition they have forgotten, review a proof they found confusing, or find the relevant section for a homework problem. Large language models (LLMs) have shown notable capabilities in domains such as mathematics~\cite{brown2020language, openai2024gpt4technicalreport}, yet they often lack alignment with specific course content and may produce hallucinated sources~\cite{ji2023survey, chen2024systematic}. For instance, prior work has shown that LLMs tend to overestimate question quality without precise course alignment~\cite{moore2022assessing} and struggle with complex mathematical problems in digital learning games~\cite{nguyen2023evaluating}. Retrieval-Augmented Generation (RAG) addresses these limitations by combining information retrieval with LLM generation, grounding responses in domain-specific documents~\cite{gao2024retrievalaugmentedgenerationlargelanguage, wang2024largelanguagemodelseducation}. Recent work has demonstrated RAG's effectiveness in building LLM-based tutors across various educational settings~\cite{han2024improving, mitra2024retllme, feng2024courseassistpedagogicallyappropriateai, lng2025rchat, zhao2025slideitright, li2024bringing}, though challenges remain around groundedness, user trust, and retrieval transparency~\cite{chiesurin-etal-2023-dangers, levonian2023retrievalaugmentedgenerationimprovemath}.

GraphRAG~\cite{edge2024localglobalgraphrag} extends standard RAG by constructing knowledge graphs from the corpus, capturing entity relationships and enabling more structured retrieval~\cite{chen2020review}. While research has shown GraphRAG can enhance question-answering performance in some settings~\cite{he2024g, lin2024research}, it may also introduce excessive or irrelevant content in fine-grained retrieval tasks~\cite{guo2024graphrag, zhang2025survey}. In educational contexts, it is critical that AI tutors not only provide correct answers but also precisely reference the pages students need~\cite{DAI2024100299, olney_improving_2024}, enabling verification and deeper learning~\cite{chiesurin-etal-2023-dangers, levonian2023retrievalaugmentedgenerationimprovemath}.

We address the following research question: \emph{``To what extent can an AI-based retrieval system identify the correct textbook page for a question derived from that specific page?''} We compare embedding-based RAG, a BM25 sparse-retrieval baseline, and GraphRAG on a custom dataset of 477 question-answer pairs from an undergraduate mathematics textbook~\cite{newstead2024infinite}, evaluating retrieval accuracy, generated answer quality (F1 score), and error patterns. Our contributions include: (1) a systematic comparison of seven retrieval approaches (five embedding models, BM25, and GraphRAG) for page-level educational QA; (2) an error analysis showing that most retrieval failures surface pedagogically relevant content from nearby pages; (3) evidence that BM25 remains a competitive baseline, outperforming several neural embedding models; and (4) practical recommendations for deploying RAG-based AI tutors.

\section{Methods}

\subsection{Dataset}

We used the undergraduate textbook \emph{An Infinite Descent into Pure Mathematics}~\cite{newstead2024infinite} as our corpus. The textbook contains 628 pages covering topics in pure mathematics. Each page was converted from PDF to LaTeX-based Markdown via GPT Vision OCR~\cite{openaiVision, ghiriti2024exploring}. We generated one question-answer pair per page (628 total) using \texttt{gpt-4o-mini}, prompting it to produce questions representative of what students might ask during self-study, for example requesting clarification of a definition, asking how to apply a theorem, or seeking the proof of a result. Two authors, who completed undergraduate coursework using this textbook, then manually reviewed all pairs, filtering out front/back matter (reducing to 528) and eliminating poorly formed or educationally irrelevant questions, yielding a final set of 477 curated pairs that reflect the types of page-specific queries students encounter when studying mathematics independently. Each pair is labeled with its source page number, enabling precise evaluation of page-level retrieval.

\subsection{RAG Pipeline}

\begin{figure*}[t]
\centering
\includegraphics[width=2\columnwidth]{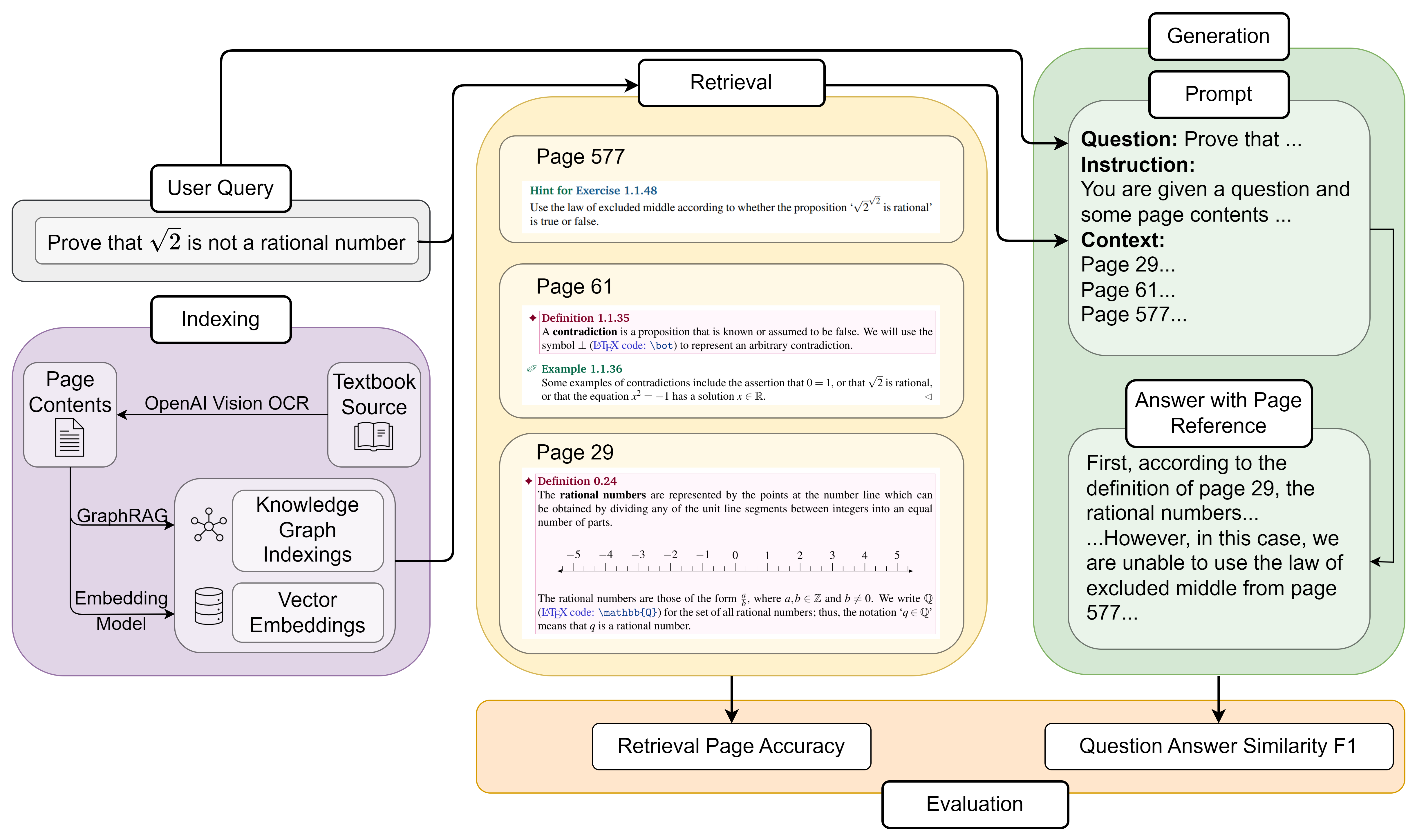}
\Description{Diagram of the RAG pipeline showing three stages: Indexing (documents encoded as vectors or graph entities), Retrieval (top-k pages selected by semantic similarity), and Generation (LLM produces answer from query and retrieved pages). Two evaluation metrics are shown: Retrieval Page Accuracy and F1 Score.}
\caption{Our RAG pipeline: Indexing, Retrieval, and Generation stages with two evaluation metrics.}
\label{fig:pipeline}
\end{figure*}

Figure~\ref{fig:pipeline} illustrates our pipeline, consisting of three stages: (1)~\textit{Indexing}, where pages are embedded as vectors (RAG), indexed as term-frequency vectors (BM25), or encoded as relational entities (GraphRAG); (2)~\textit{Retrieval}, where the top-$k$ most relevant pages or entities are identified; and (3)~\textit{Generation}, where the query, prompt, and retrieved content are fed to an LLM for answer generation.

\subsection{Models and Baselines}

We selected five embedding models from the December 2024 Massive Text Embedding Benchmark (MTEB) leaderboard~\cite{muennighoff2022mteb, huggingfaceMTEBLeaderboard}: \texttt{voyage-3-large} and \texttt{nv-embed-v2} were chosen for their high rankings; \texttt{gte-large} is commonly used as a fine-tuning base; \texttt{text-embedding-3-large} was included for its widespread adoption; and \texttt{multilingual-e5} for its multilingual capabilities. We also include a \textbf{BM25} sparse-retrieval baseline using Okapi BM25 with default parameters, to contextualize neural embedding gains against a classical term-matching approach that requires no training or fine-tuning. For each model, we tested retrieval at $k \in \{1, 3, 5, 10\}$.

For GraphRAG~\cite{edge2024localglobalgraphrag}, we adapted the framework to expose \texttt{document\_ids} for each entity and text unit, enabling page-level traceability. Specifically, we modified various GraphRAG components to store and output fields such as \texttt{document\_ids} and \texttt{entity\_ids}, allowing our system to link each retrieved item back to a corresponding page number. We tested GraphRAG with both \texttt{gpt-4o-mini} and \texttt{o3-mini} as the underlying LLM. Note that since GraphRAG retrieves entire entities and community summaries rather than ranked pages, we cannot directly control the number of pages returned; accuracy is evaluated based on whether the correct page's entities appear in the retrieved data.

\subsection{Evaluation Metrics}

We evaluate with two metrics: (1)~\textbf{Retrieval Accuracy}, whether the correct source page appears in the retrieved set, averaged across all 477 queries; and (2)~\textbf{F1 Score}, word-overlap F1 between the generated answer (produced by \texttt{gpt-4o-mini} given retrieved context) and the ground-truth answer. The F1 score captures the balance between precision (fraction of generated words appearing in the reference) and recall (fraction of reference words appearing in the generated output). We report bootstrap 95\% confidence intervals (10,000 resamples) for key comparisons to assess statistical significance. We also conducted an error analysis of retrieval failures, categorizing top-1 misses by page distance and chapter membership to understand whether failures are near-misses or far-misses. Additionally, we tested LLM-based re-ranking of retrieved pages using \texttt{gpt-4o-mini}, where the model reorders retrieved pages by estimated relevance to the query.

\section{Results}

\subsection{Retrieval Accuracy}

Table~\ref{tab:retrieval} presents retrieval accuracy results. Among all models, \texttt{voyage-3-large} achieves the highest accuracy across all top-$k$ settings, reaching 99.4\% at top-10. Bootstrap 95\% CIs for top-1 accuracy confirm that \texttt{voyage-3-large} ([.644, .728]) significantly outperforms \texttt{nv-embed-v2} ([.541, .629]), as these intervals do not overlap at top-1 or top-3.

BM25 proves a strong sparse-retrieval baseline, achieving top-1 accuracy of .579, outperforming both \texttt{gte-large} (.461) and \texttt{multilingual-e5} (.457) at all $k$ values despite using no learned representations. This highlights the importance of including classical baselines in retrieval evaluations. Despite strong MTEB leaderboard rankings, \texttt{multilingual-e5} underperformed on this domain-specific task, underscoring that benchmark performance does not always transfer to specialized mathematical domains.

GraphRAG achieves 84.5\% with \texttt{gpt-\allowbreak 4o-\allowbreak mini}
and 91.4\% with \texttt{o3-\allowbreak mini}. All five embedding models
surpass GraphRAG's accuracy at top-\allowbreak 5 or above. The gap between
the two GraphRAG configurations suggests that the underlying LLM quality
substantially affects graph-based retrieval performance.

\begin{table}[t]
\caption{Retrieval accuracy for embedding-based RAG models, BM25, and GraphRAG. GraphRAG retrieves entities rather than ranked pages, so only overall accuracy is reported.}
\centering
\small
\begin{tabular}{lcccc}
\toprule
\textbf{Model} & \textbf{Top-1} & \textbf{Top-3} & \textbf{Top-5} & \textbf{Top-10} \\
\midrule
voyage-3-large & \textbf{.686} & \textbf{.910} & \textbf{.958} & \textbf{.994} \\
nvidia/nv-embed-v2 & .585 & .843 & .912 & .964 \\
BM25 & .579 & .780 & .855 & .929 \\
OpenAI emb-3-large & .549 & .811 & .893 & .933 \\
gte-large & .461 & .711 & .799 & .881 \\
multilingual-e5 & .457 & .702 & .795 & .870 \\
\midrule
GraphRAG (o3-mini) & \multicolumn{4}{c}{.914} \\
GraphRAG (4o-mini) & \multicolumn{4}{c}{.845} \\
\bottomrule
\end{tabular}
\label{tab:retrieval}
\end{table}

\subsection{Generated Answer Quality}

Table~\ref{tab:f1} shows F1 scores for answer generation. The no-retrieval baseline (\texttt{gpt-4o-mini} alone) achieves 0.475. All RAG configurations improve upon this, with F1 scores ranging from 0.514 to 0.552, confirming that retrieval grounding consistently benefits domain-specific QA. GraphRAG achieves F1 of $\sim$0.524, comparable to but slightly below the best embedding-based results.

Interestingly, \texttt{nv-embed-v2} achieves the highest top-1 F1 (.537) despite lower retrieval accuracy than \texttt{voyage-3-large}, suggesting retrieval accuracy and generation quality do not always correlate, because near-miss retrievals still provide useful context. Increasing $k$ generally does not improve F1, as additional pages introduce noise. GraphRAG's $\sim$47K tokens of context per query (vs.\ $\sim$3.7K for top-5 RAG) overwhelms the generator, suggesting that \emph{retrieval precision} matters more than \emph{recall} for generation quality.

\begin{table}[t]
\caption{F1 scores for generated answers. Baseline is \texttt{gpt-4o-mini} without retrieval (F1 = 0.475).}
\centering
\small
\begin{tabular}{lcccc}
\toprule
\textbf{Model} & \textbf{Top-1} & \textbf{Top-3} & \textbf{Top-5} & \textbf{Top-10} \\
\midrule
voyage-3-large & .523 & .543 & .544 & .547 \\
nvidia/nv-embed-v2 & \textbf{.537} & .542 & .541 & .539 \\
OpenAI emb-3-large & .531 & .549 & .533 & .543 \\
gte-large & .526 & .533 & .550 & \textbf{.552} \\
multilingual-e5 & .514 & .534 & .541 & .535 \\
\midrule
GraphRAG (o3-mini) & \multicolumn{4}{c}{.524} \\
GraphRAG (4o-mini) & \multicolumn{4}{c}{.525} \\
\midrule
No retrieval & \multicolumn{4}{c}{.475} \\
\bottomrule
\end{tabular}
\label{tab:f1}
\end{table}

\subsection{Re-Ranking}

LLM-based re-ranking with \texttt{gpt-4o-mini} yielded mixed results. Re-ranking top-5 pages degraded \texttt{voyage-3-large}'s top-1 accuracy from .686 to .593, while re-ranking top-10 caused further degradation across all models. Critically, we observed hallucinated page numbers in re-ranking outputs: the LLM occasionally referenced pages not in the retrieved set. These findings suggest that LLM-based re-ranking is unreliable for page-level retrieval, particularly with larger candidate sets.

\subsection{Error Analysis}

We examined the 150 top-1 failures from \texttt{voyage-3-large}, categorizing each by page distance and chapter membership. Of these, 46.0\% retrieved an adjacent page (distance $\leq$ 2), 17.3\% retrieved a same-chapter but non-adjacent page, and 35.3\% were far misses (different chapter). Overall, 63.3\% of failures retrieved same-chapter content, with a median page distance of just 3 pages. This suggests that even ``incorrect'' retrievals often surface pedagogically relevant material. For example, a student asking about a proof may be directed to the adjacent page containing the theorem statement, content that is still valuable for building understanding and that an instructor might recommend reviewing first. The high proportion of near-misses reflects the textbook's pedagogical structure, where definitions, theorems, and proofs are deliberately organized across consecutive pages to support incremental learning. The 35.3\% far misses, where shared terminology across chapters causes cross-topic confusion, represent the genuinely problematic failures that could misdirect students.

\subsection{Open-Source LLM Replication}

To assess whether our findings generalize beyond commercial APIs, we replicated key experiments using Qwen3.5-35B-A3B~\cite{qwen3.5}, an open-source 36B Mixture-of-Experts model running locally on 2$\times$ NVIDIA TITAN RTX GPUs with Q4\_K\_M quantization via Ollama. For retrieval, we tested \texttt{nomic-embed-text}~\cite{nussbaum2024nomic}, an open-source embedding model, also served locally.

Table~\ref{tab:opensource} compares generation quality (F1) between the commercial \texttt{gpt-4o-mini} and the local Qwen3.5 setup. Without retrieval, Qwen3.5 achieves substantially lower F1 (0.276 vs.\ 0.475). However, adding RAG with \texttt{nomic-embed-text} retrieval improves Qwen3.5's F1 to 0.384, a \textbf{39\% relative improvement}, compared to only 16\% for \texttt{gpt-4o-mini} with its best embedding model. This suggests that \emph{weaker models benefit disproportionately from retrieval grounding}, as they rely more heavily on external context to compensate for limited parametric knowledge.

\begin{table}[t]
\caption{F1 scores: commercial API vs.\ open-source local LLM.}
\centering
\small
\begin{tabular}{lcc}
\toprule
\textbf{Configuration} & \textbf{gpt-4o-mini} & \textbf{Qwen3.5-35B} \\
\midrule
No retrieval & 0.475 & 0.276 \\
Best RAG + generation & 0.552 & 0.384 \\
Relative improvement & +16\% & +39\% \\
\bottomrule
\end{tabular}
\label{tab:opensource}
\end{table}

For retrieval accuracy, \texttt{nomic-embed-text} achieved top-1 accuracy of 45.1\%, top-3 of 71.3\%, and top-10 of 88.7\%, below the best commercial embeddings but still serviceable. Notably, BM25 (61.2\% top-1) again outperforms this neural embedding model, reinforcing BM25's strength as a zero-cost baseline.

We also attempted GraphRAG v3.0.6 indexing with Qwen3.5, successfully extracting 4{,}528 entities, 10{,}828 relationships, and 1{,}107 communities. However, query-time generation failed due to incompatibilities between GraphRAG's OpenAI-compatible API expectations and Qwen3.5's thinking-mode architecture. This demonstrates that while local LLM indexing for GraphRAG is feasible, end-to-end integration with open-source models remains challenging.

\section{Discussion and Conclusion}

Our findings yield several key insights for designing RAG-based educational AI systems.

\textbf{RAG consistently improves over baseline LLMs.} All RAG configurations produced higher F1 scores than the no-retrieval baseline (0.475 vs.\ 0.514 to 0.552), confirming that grounding LLM responses in textbook content improves answer quality for domain-specific question answering. This improvement is meaningful in educational settings where students benefit from answers grounded in their assigned course materials rather than the LLM's general pre-training data, which may contain incorrect or outdated mathematical content~\cite{chen2025mathproof_arxiv, chen2026chatbased}.

\textbf{Embedding-based RAG outperforms GraphRAG for page-level retrieval.} GraphRAG retrieves excessive context ($\sim$47K tokens vs.\ $\sim$3.7K for top-5 RAG), leading to lower precision and comparable or lower F1~\cite{guo2024graphrag, zhang2025survey}. This reflects a \emph{task-method mismatch}: GraphRAG was designed for global summarization and multi-hop reasoning, not page-level lookup. For educational applications where students need a specific page reference, the simpler embedding-based approach is more appropriate. However, GraphRAG may excel at cross-concept questions (e.g., ``How do induction and well-ordering relate across chapters?'') where its entity-based structure could surface connections that page-level retrieval would miss.

\textbf{BM25 is a competitive baseline.} BM25 outperforms two of five embedding models at all $k$ values, partly due to high lexical overlap (72\% unigram) between questions and source pages. This underscores the importance of including classical baselines, as neural embeddings do not always yield proportional gains, particularly when questions and documents share substantial vocabulary.

\textbf{Most retrieval failures are near-misses with pedagogical value.} Our error analysis reveals that 63.3\% of top-1 failures retrieve same-chapter content (median distance: 3 pages). From an educational perspective, this means a student directed to the ``wrong'' page is still reading closely related material in most cases, analogous to how a tutor might say ``look at the previous page first'' before addressing the specific question. The top-1 accuracy of 68.6\% may appear insufficient, but the near-miss pattern means the system rarely sends students to entirely irrelevant content.

\textbf{Re-ranking introduces more risk than benefit.} LLM-based re-ranking degraded performance for top-performing retrievers and introduced hallucinated page references, suggesting general-purpose LLMs struggle to distinguish between mathematically similar pages. Deployments requiring re-ranking should explore fine-tuned cross-encoder models or constrained decoding.

\textbf{Open-source LLMs are viable but benefit more from RAG.} RAG provides proportionally larger gains for weaker models (+39\% vs.\ +16\% relative F1 improvement), suggesting that retrieval grounding partially compensates for reduced model capability. This is significant for educational settings where commercial API costs are prohibitive, and aligns with EDM 2026's ``Across Borders'' theme, because institutions can deploy useful textbook-grounded QA systems on local hardware rather than relying on commercial APIs. Our results show that a locally hosted open-source model combined with BM25 retrieval (requiring no GPU for the retrieval component) can deliver meaningful QA quality at near-zero marginal cost.

\textbf{Practical recommendations.} (1)~Use top-5 RAG retrieval as default (accuracy .958, $\sim$3.7K tokens); (2)~present top-3 candidate pages to students (accuracy $>$91\%); (3)~avoid LLM re-ranking due to hallucination risk; (4)~consider adjacent-page windowing to capture theorem-proof pairs spanning page boundaries.

\textbf{Limitations and Future Work.} While questions were curated by course-familiar authors to reflect realistic student queries, their high lexical overlap with source pages (72\% unigram) may inflate retrieval accuracy relative to more diverse student phrasings; future work should supplement with authentic questions from discussion forums or office hours. Word-overlap F1 is a weak proxy for math QA, as mathematically equivalent expressions can score poorly; LLM-as-judge or symbolic equivalence checking would better capture correctness. Extending to multiple STEM textbooks, investigating hybrid BM25+dense retrieval, and conducting classroom studies measuring whether page references improve comprehension and student trust are important next steps. Notably, \texttt{multilingual-e5}'s underperformance despite strong general benchmarks raises questions about deploying these systems for non-English textbooks, an important consideration for educational equity across linguistic borders.

\bibliographystyle{abbrv}
\bibliography{main}

\end{document}